\documentclass[twocolumn,prb,amsmath,showpacs,amssymb,citeautoscript,floatfix]{revtex4}
\def\etal{{ \it et al. }}
\def\prb{{Phys. Rev. B}}
\def\pl{{Phys. Rev. Lett.}}
\usepackage{epsfig}
\newcommand{\rb}[1]{\raisebox{1.5ex}[-1.5ex]{#1}}  
\begin{document}
\title{ Linear response
results for phonons and electron-phonon coupling in hcp Sc - spin fluctuations and implications for superconductivity} 
\author{S. K. Bose}
\affiliation{Physics Department, Brock University, St. Catharines,
Ontario L2S 3A1, CANADA                    } 
\begin{abstract}
We present a FP-LMTO (full potential linear muffin-tin orbitals) study of the variation in
the electronic structure, phonon frequencies and electron-phonon coupling in hcp Sc under pressure.
   The electron-phonon coupling constant $\lambda$ is found to
increase steadily with pressure in the hcp phase, until the
 pressure reaches a value where the hcp phase becomes unstable. Linear response calculations for the normal
pressure $c/a$ ratio  predict a phase change somewhere between calculated pressures of 22 and 30 GPa. 
The calculated frequencies for the equilibrium hcp lattice parameters are in good agreement with the
inelastic neutron scattering results. There is a small upward shift in the
$\Gamma$-point $E_{2g}$ mode frequency under pressure, in qualitative agreement with the Raman spectroscopy study of Olijnyk \etal (J. Phys.:Condens. Matter,{\bf 18}, 10971 (2006)). From the measured value of the electronic specific heat constant and the calculated values of the Fermi level density of states and electron-phonon coupling constant, we conclude that the electron-paramagnon coupling constant in hcp Sc should be comparable to the electron-phonon coupling constant. This indicates that the spin fluctuation effects are strong enough to suppress superconductivity completely in hcp Sc. We argue that spin fluctuations should be reduced by a factor of two or more in the high pressure Sc-II phase. Based on estimates of the electron-paramagnon coupling constants and the calculated or estimated electron-phonon coupling constants, we argue that the hcp phase may become superconducting with a very low transition temperature immediately prior to the transition to the Sc-II phase and that the Sc-II phase should indeed be superconducting. The electronic, electron-phonon and spin fluctuation properties in hcp Sc under pressure are compared with those of the high pressure hcp phase of Fe, which was  reported to be superconducting a few years back. 
 
\end{abstract} 
\pacs{74.25.Jb, 74.25.Kc, 74.70.Ad, 74.90.+n}

\maketitle 
\section{Introduction} 

This work was motivated by a recent work of Hamlin and Schilling \cite{Schilling07}, who reported the 
measurement of the superconducting transition temperature $T_c$ in Sc as a function of pressure. 
  Superconductivity  in  Sc is induced under pressure, with $T_c$ increasing monotonically to 8.2 K at 74.2 GPa. 
Hamlin and Schilling \cite{Schilling07} report measurements of $T_c$ between $\sim 55$ GPa ($T_c\sim 5 $K) and 
74.2 GPa ($T_c= 8.2$ K).  The normal pressure phase of Sc is hcp and it is known to
undergo several changes in its crystal
structure as a function of pressure, the first such transition known to be occurring around 22-23 GPa 
\cite{Fujihisa05,Akahama05,McMahon06} from hcp (Sc-I) to a complicated structure, referred to as Sc-II.
To date the normal and low pressure hcp phase is known to be non-superconducting. Earlier work by Wittig \etal \cite{Wittig79} showed that superconductivity is possibly induced in the hcp phase at high pressure around 20 GPa, immediately before it enters the complex Sc-II phase. $T_c$ in the hcp phase was estimated to be less than 0.1 K.

  {\it Ab initio} theoretical studies of superconductivity
in Sc as a function of pressure is rendered difficult by the fact that the superconducting Sc-II phase is not only complex, but also that its exact structure still remains open to  investigation and refinement \cite{Fujihisa05,Akahama05,McMahon06}.
 Ormeci \etal \cite{Ormeci06} have studied the electronic structure of the
Sc-II phase proposed by Fujihisa \etal \cite{Fujihisa05} and McMahon \etal \cite{McMahon06} and produced results suggesting that the structure proposed by McMahon \etal might provide a better representation of the Sc-II phase. Both of these proposed structures are 
composite incommensurate structures, consisting of host and guest substructures.

In this communication we present {\it ab initio} calculations of the electronic structure, elastic properties,
phonons and electron-phonon coupling in the low pressure hcp phase of Sc. The purpose is to shed some light on
why hcp Sc is not superconducting, while the substance might become superconducting under pressure. Several scenarios are possible. It might be that the electron-phonon coupling in the hcp phase is small, making superconductivity unlikely in this phase. This coupling may increase with pressure, but still remain too small
for superconductivity to appear over the entire pressure range for which the hcp phase is stable. Another possibility is that the electron-phonon coupling is strong enough to support superconductivity in the hcp phase.
In this latter scenario one needs to explore why superconductivity is suppressed in the hcp phase and appears only in the high pressure Sc-II phase. Calculations of the phonon frequencies as a function of pressure should also show  softening of certain modes leading to the instability of the hcp phase at high pressure. 

We use the full potential linear muffin-tin orbital (FP-LMTO)
method \cite{savrasov-el} to study the electronic structure and the linear response scheme developed by Savrasov \cite{savrasov1,savrasov2} to compute the phonon-frequencies, the Eliashberg spectral function and the 
electron-phonon coupling constant $\lambda$ as a function of volume in hcp Sc. For the sake of simplicity and
convenience in the calculation, the $c/a$ ratio is kept fixed at the normal pressure value \cite{Kittel}. 
This restriction of the $c/a$ ratio being kept fixed at the normal pressure value should not be of any concern from the viewpoint of the main results and conclusions of this work. This will become clear from the discussion in the following sections.

Sc is the lightest of the $3d$- transition metals. It would be of interest to see how the electron-phonon coupling
in hcp Sc compares with that of the late transition metal hcp Fe, which was reported to be superconducting by Shimizu\etal \cite{nature1} (see also Ref.\onlinecite{jaccard}) a few years back. The hcp phase is the stable phase of Fe at pressures $\sim$ 10 GPa and higher. Bose\etal \cite{Bose03} reported FP-LMTO linear-response results for hcp Fe. They showed, in agreement with  
earlier LAPW(linear augmented plane-wave)-based rigid muffin-tin (RMT) results of Mazin and coworkers \cite{mazin1}, that not only should hcp Fe be superconducting, but also conventional (electron-phonon)  $s$-wave superconductivity in hcp Fe should persist up to a much higher pressure than what is found in the experiments. Both ferromagnetic and antiferromagnetic spin fluctuation effects were considered in the works of Bose\etal \cite{Bose03} and Mazin\etal \cite{mazin1}  The conclusion was that such spin fluctuation effects in hcp Fe could lower $T_c$ somewhat, but could not account for the rapid disappearance of $T_c$
with increasing pressure, as observed by Shimizu \etal \cite{nature1} Nontrivial differences between the electronic structures of Fe and Sc, due in particular to the large difference in the number of $d$-electrons, cause significant differences in their elastic as well as electron-phonon scattering properties, which should have opposite effects on the overall electron-phonon coupling constant.

Ormeci \etal \cite{Ormeci06} have used crystalline approximants of the incommensurate structures proposed by Fujihisa \etal \cite{Fujihisa05} and McMahon \etal \cite{McMahon06} in studying the electronic structure of the Sc-II phase. The unit cells of these approximants contain 22 atoms for the model proposed by 
Fujihisa \etal \cite{Fujihisa05} and 42 atoms for the model suggested by McMahon \etal \cite{McMahon06}.
Because of the high computational demand due to the large number of atoms in the unit cells of these approximants, we are unable to extend  the linear response calculations of phonons and electron-phonon coupling to the Sc-II phase.

\section{Electronic structure}

\begin{figure}
\resizebox{!}{2.5in}{\includegraphics{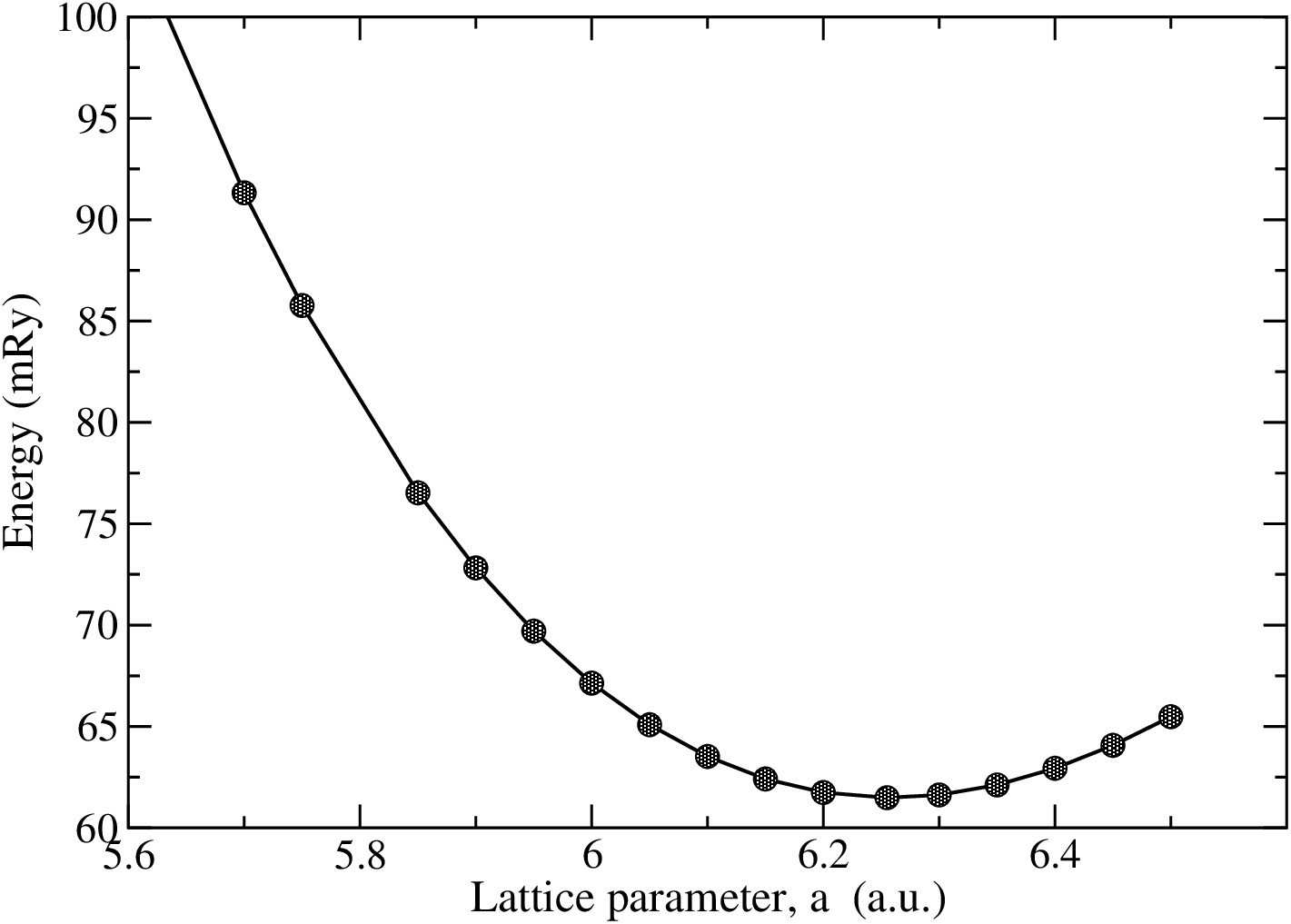}}
\caption[]{ FP-LMTO total energy per atom versus lattice parameter in hcp Sc for the $c/a = 1.592$,
valid for the normal pressure values: $a=3.31$ \AA, $c=5.27$ \AA [\onlinecite{Kittel}]. For convenience in plotting,
a constant value of -1529.0 Ry from the total energy per atom has been subtracted.} 
\label{fig1}
\end{figure}

FP-LMTO results of total energy as a function of the lattice parameter in hcp Sc is shown in Fig.\ref{fig1}. The $c/a$ ratio was kept fixed at the normal pressure value \cite{Kittel}. As in the earlier calculation for hcp Fe \cite{Bose03}, the generalized gradient approximation of Perdew and Wang (GGA1)\cite{PW1} for the exchange-correlation potential was used.  The electronic structure was computed using a two-$\kappa$ $spd$ LMTO basis for the valence band. 3$s$- and 3$p$-semi-core states were treated as 
valence states in separate energy windows.  The charge densities and potentials were represented by spherical harmonics with $l\leq 6$ inside the 
non-overlapping MT spheres and by plane waves with energies $\leq$ 48-70 Ry, depending on the lattice parameter, in the interstitial region.
Brillouin zone (BZ) integrations were performed with the full-cell tetrahedron method \cite{peter}, using 1200
{\bf k}-points in the irreducible zone. The pressure and bulk modulus calculated from this energy-volume curve, using the generalized Birch-Murnaghan equation of state \cite{birch,murnaghan}, are shown in Table \ref{table1}.
 Note that the minimum energy occurs almost exactly at the normal pressure lattice parameter of 6.255 a.u. The calculated bulk modulus of 56 GPa at this \cite{Kittel} equilibrium lattice parameter is somewhat higher than the
experimental value, 43.5 GPa.  One interesting observation is that the lattice parameter 5.75 a.u., or the volume per atom $V_0 = 131$ a.u., corresponds to a pressure of 22 GPa. This is the pressure at which the hcp Sc-I phase becomes unstable against the formation of a complicated structure known as the Sc-II phase \cite{Fujihisa05,Akahama05,McMahon06}. As will be discussed later, the linear response results show that the phonon frequencies calculated for lattice parameter $a=$ 5.6 a.u. are imaginary. Thus, for the normal pressure $c/a$ ratio, the hcp structure becomes unstable at some lattice parameter between 5.75 and 5.6 a.u. The calculated pressure at 5.6 a.u. is $\sim$ 30 GPa. Given the uncertainties in the pressure calculation, this result is in good agreement with the experimental observation of a phase change in Sc at 22-23 GPa. Additional linear response calculations to better locate the phase change were not done.

\begin{table}
\caption{FP-LMTO results for  hcp Sc for the normal pressure $c/a$ ratio, 1.592. a= lattice parameter (a.u.),
$V_0$ = volume per atom (a.u.), $P$=pressure (GPa), $B$=Bulk Modulus (GPa), $N(0)$= DOS at the Fermi level
(states/(Ry cell)).}
\label{table1}
\begin{ruledtabular}
\begin{tabular}{lcccccc}
a  & 6.255 & 6.15& 6.05 & 5.95& 5.85& 5.75\\
$V_0$  & 168.72 & 160.37 & 152.67 & 145.22 & 138.02 & 131.07\\
$P$  & -0.084 & 3.21 & 6.85 & 11.22 & 16.20 & 22.28 \\
$B$  & 56.2 & 67.2 & 79.6 & 93.0 & 108 &  125\\
$N(0)$  & 58.06 & 54.82  & 51.59  & 49.07  & 47.27  & 47.71 \\
\end{tabular}
\end{ruledtabular}
\end{table}

\begin{figure}
\resizebox{!}{2.5in}{\includegraphics{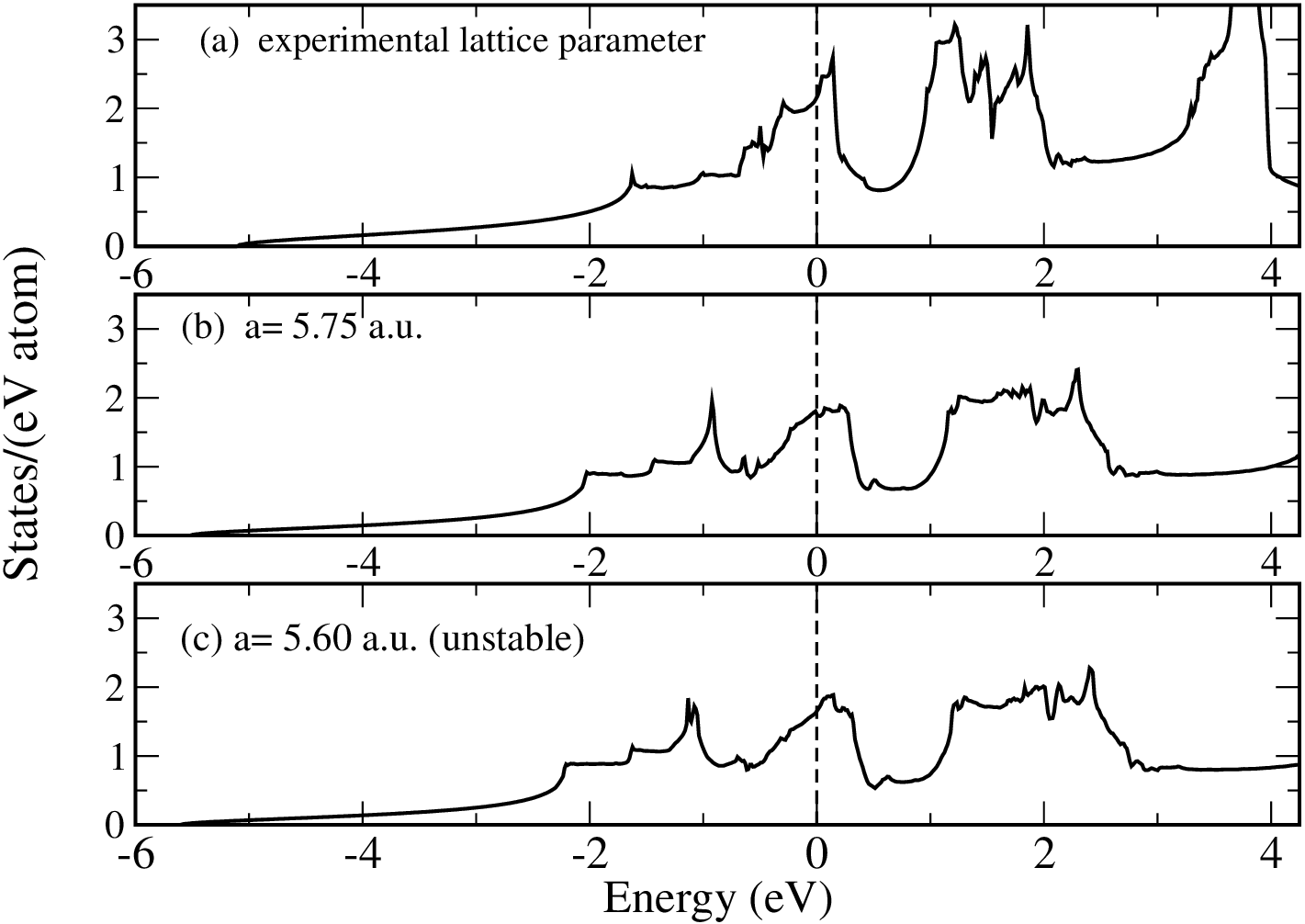}} 
\caption[]{ FP-LMTO DOSs for three lattice parameters of hcp Sc: (a) Normal pressure values[\onlinecite{Kittel}]: $a = 6.255$ a.u., $c/a=1.592$, (b) $a=5.75$ a.u. and (c) $a=5.60 a.u.$; $c/a$ ratio for (b) and (c) is the same as in (a). It appears that as the Fermi level moves from the shallow valley region between two small peaks (a) to a shoulder (c),  the hcp structure becomes unstable.}
\label{fig2}
\end{figure}

The FP-LMTO DOS calculated for three lattice parameters are shown in Fig.\ref{fig2}. Panel (a) shows the DOS for the experimental normal pressure lattice parameters: $a=3.31$ \AA, $c=5.27$ \AA \cite{Kittel}. Panels (b) and (c) show the
DOSs for the lattice parameters 5.75 a.u. and 5.60 a.u., respectively, with the same $c/a$ ratio as in (a). For the equilibrium lattice parameters (panel (a)), the Fermi level falls in a shallow valley between two small peaks.
With increasing pressure, the bands broaden, flattening these peaks and the valley region (panel (b)). It appears that as this region around the Fermi level changes to a shoulder (panel (c)), the hcp phase becomes unstable.

Hamlin and Schilling \cite{Schilling07} discuss the importance of the role of $s \rightarrow d$ charge transfer 
 in the variation of electron-phonon coupling and the changes in the crystal structure under pressure.
 Pettifor \cite{Pettifor77} has discussed the variation of the $d$-band occupancy under pressure of $4d$- transition metals, showing that the increase in the $d$-electron occupancy of the early transition metals under pressure is rapid.  Duthie and Pettifor \cite{Pett-Duth77} used a $s-d$ band model to  discuss the correlation between 
the $d$-band occupancy and the crystal structures in rare earths. They showed that $d$-electron concentration
 $n_d$ increases under pressure and a sequence of crystal structure changes takes place as $n_d$ changes from 1.5 to 2.5. With increasing pressure, the volume available to the electrons outside the ion cores  diminishes
rapidly. As a result, simple metals like Li and Na are known to become non-free-electron-like under pressure and Cs is known to become a transition metal as the $5d$-band begins to fill due to $s\rightarrow d$ charge transfer (see references in [\onlinecite{Schilling07}]). In any electronic structure calculation the numbers of $s$-, $p$- and $d$- electrons are essentially functions of the basis set used, if  such a division  is at all possible for the method used. It is easier to keep track of $s\rightarrow d$ charge transfer in LMTO-ASA (atomic sphere approximation)
\cite{varenna}, where the basis consists of muffin-tin orbitals only, rather than in FP-LMTO. In Table \ref{table2} we
present the orbital- or partial wave channel-resolved electron numbers as a function of the lattice parameter, as obtained by the LMTO-ASA method \cite{footnote}. The results presented were obtained by using the exchange correlation potential of Perdew and Wang \cite{PW92} in the local density approximation. Checks for a couple of lattice parameters using GGA1 \cite{PW1} had revealed similar results. The decrease in $n_s$ and the increase in $n_d$ are monotonic
as a function of decreasing lattice parameter. In fact, charge transfer takes place from the extended $s$- and $p$-orbitals to the less extended or localized $d$- and $f$-orbitals. Of particular interest are  $n'$ and $n^-$, defined as
$n'=\frac{n_d+n_f}{n_s+n_p}$ and $n^- = (n_d+n_f)-(n_s+n_p)$, quantifying the transfer of electrons from the delocalized to the localized channels. At normal pressure the electrons are divided almost equally between the
localized and delocalized channel ($n^-=0.15$).  With increasing pressure the electrons  more and more
 occupy the localized orbitals.
\begin{table}
\caption{LMTO-ASA results for  (approximate) $s$-, $p$-, $d$- and $f$-orbital resolved charges: $n_s, n_p, n_d, n_f$ and the corresponding Fermi level DOSs: $N_s, N_p, N_d, N_f$. $n'=\frac{n_d+n_f}{n_s+n_p}$, 
$n^- = (n_d+n_f)-(n_s+n_p)$ and $N'=
\frac{N_d+N_f}{N_s+N_p}$ are presented to reveal the trends in the inter-orbital charge transfer and the redistribution of the orbital-resolved DOSs as a function of volume per atom. $N(0)$ is the total DOS at the Fermi level. All DOS are in units of states/(Ry cell). a= lattice parameter (a.u.)}
\label{table2}
\begin{ruledtabular}
\begin{tabular}{lcccccc}
a  & 6.255 & 6.15& 6.05 & 5.95& 5.85& 5.75\\
$n_s$  & 0.730  & 0.720  & 0.714  & 0.704  & 0.695  & 0.685 \\
$n_p$  & 0.690 & 0.677& 0.662& 0.645 & 0.626 & 0.603 \\
$n_d$  & 1.545 & 1.567 & 1.589 & 1.614 & 1.641 & 1.672\\
$n_f$  & 0.031 & 0.033 & 0.034 & 0.036 & 0.037 & 0.039\\
$n^-$  & 0.156 & 0.203 & 0.247 & 0.301 & 0.357 & 0.423\\
$n'$ & 1.11 & 1.15 & 1.18 & 1.22 & 1.27 & 1.33\\
\hline
$N(0)$ & 57.89 & 54.78 & 53.03 & 51.60 & 50.16 & 53.34\\
$N_s(0)$ & 0.941 & 0.946 & 0.964 & 0.982 & 1.020 & 1.226\\
$N_p(0)$ & 14.836 & 13.729 & 12.889 & 12.257 & 11.914 & 13.558\\
$N_d(0)$ & 40.789 & 39.110 & 37.887 & 37.080 & 36.444 & 37.222\\
$N_f(0)$ & 1.326 & 1.301 & 1.287 & 1.280 & 1.280 & 1.335\\
$N'$ & 2.67 & 2.75 & 2.83 & 2.90 & 2.92 & 2.60\\
\end{tabular}
\end{ruledtabular}
\end{table}

As the pressure increases different partial bands broaden at different rates. In Table \ref{table2} we also show the changes in partial or angular momentum-resolved DOSs at the Fermi level. The ratio of the localized orbital DOSs ($N_d(0)$+
$N_f(0)$) to the delocalized orbital DOSs ($N_s(0)$+$N_p(0)$) seem to increase steadily under pressure, except for the lowest lattice parameter shown. The decrease in $N'$ for the lattice parameter 5.75 a.u. could be
specific to LMTO-ASA, as the somewhat large change  in $N(0)$ between the lattice parameters 5.85 and 5.75 a.u. is not observed in the FP-LMTO results (Table \ref{table1}).

Ormeci \etal \cite{Ormeci06} have calculated the partial charges in the high pressure Sc-II phase. They use  the
full-potential local-orbital (FPLO) method of Koepernik and Eschrig\cite{Eschrig} with an $spd$-basis. At 
ambient pressure, in the hcp structure, their $nl$-projected charges for the $4s$, $4p$, and $3d$ channels are
0.70, 0.59 and 1.70, respectively. In the Sc-II phase at about  two thirds of the ambient pressure volume per atom, these numbers become 0.54-0.65, 0.27-0.51, and 1.92-2.05, respectively. Their calculations also suggest that
a complete $s\rightarrow d$ charge transfer would require pressures in excess of 240 GPa.

\section{Phonons and electron-phonon coupling}
\begin{figure*}
\resizebox{!}{5.0in}{\includegraphics{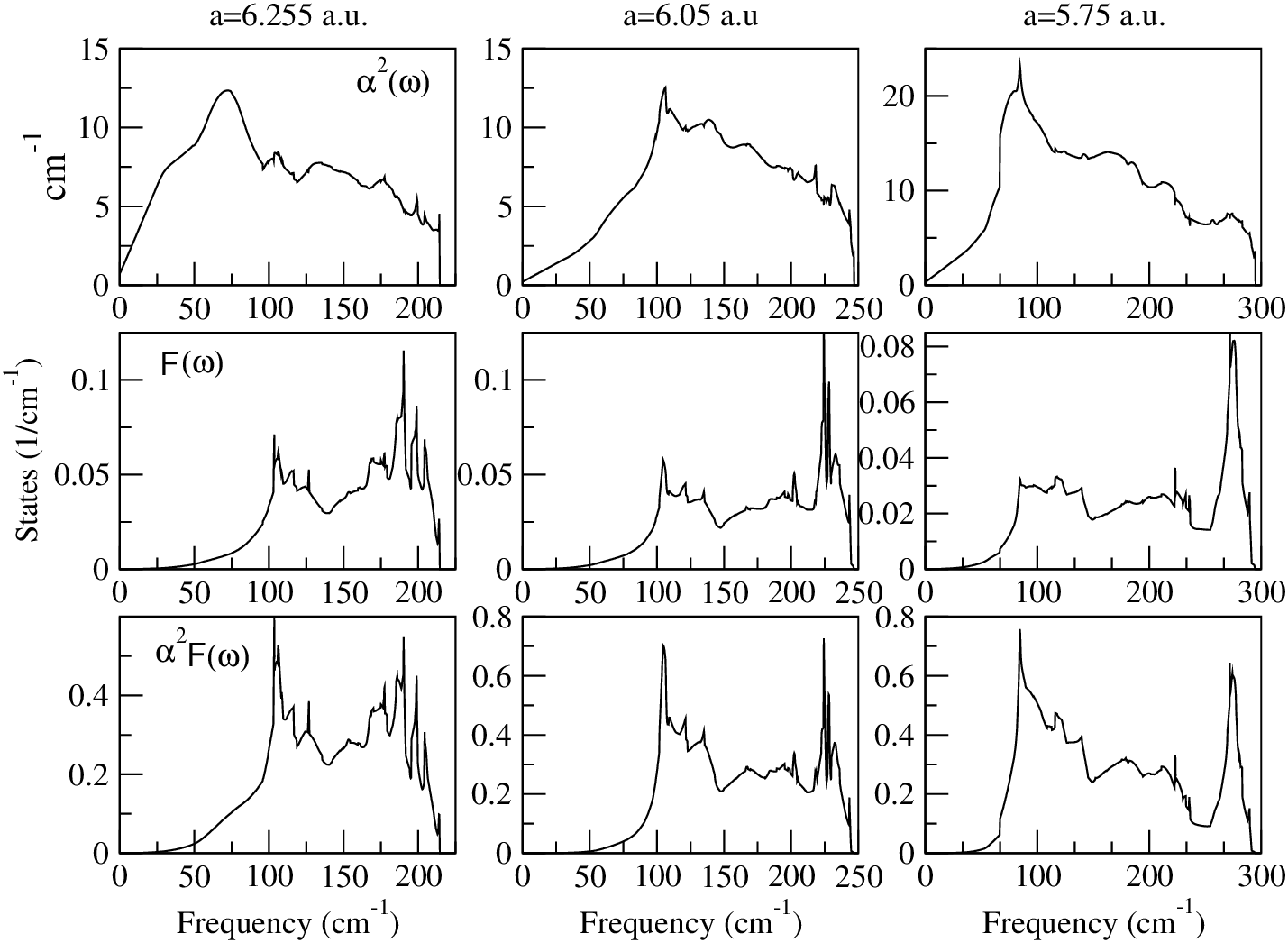}}
\caption[]{Phonon density of states $F(\omega)$, the Eliashberg spectral
function $\alpha^2F(\omega)$ and the function $\alpha^2(\omega)$, defined
as the ratio $\alpha^2F(\omega)/F(\omega)$, for three lattice parameters $a=$ 6.255 (experimental normal pressure value), 6.05, and 5.75 a.u.,
with $c/a$ ratio fixed at the normal pressure value. }
\label{fig3}
\end{figure*}

The FP-LMTO linear response results for phonons and electron-phonon coupling are summarized in Table \ref{table3}.
 The dynamical matrix  was generated for 32 phonon wave vectors in the irreducible BZ, corresponding to a mesh of (7,7,7) reciprocal lattice divisions.  The BZ sums  for the dynamical and electron-phonon (Hopfield) matrices was done for a (28,28,28) mesh, resulting in 1200 wave vectors in the IBZ.  These choices were based on extensive tests performed  by Savrasov and Savrasov\cite{savrasov1,savrasov2} in their study of the phonons and electron-phonon coupling in elemental metals and alloys. Phonon density of states $F(\omega)$, Eliashberg spectral function 
$\alpha^2F(\omega)$ and the function $\alpha^2(\omega) = \alpha^2F(\omega)/F(\omega)$ are shown in Fig. \ref{fig3}
for three different lattice parameters with the $c/a$ ratio fixed at the normal pressure value.

The calculated phonon frequencies appear to be in reasonably good agreement with the inelastic neutron 
scattering (INS) results\cite{Wakabayashi71,Reichardt}. For the normal pressure lattice parameter (6.255 a.u.), the calculated longitudinal and transverse optic (LO and TO) mode frequencies at the $\Gamma$-point are
 6.273 THz (209 cm$^{-1}$) and 3.912 THz (130.5 cm$^{-1}$), respectively. These compare well with the
inelastic neutron scattering results of Wakabayashi \etal \cite{Wakabayashi71}: 6.91 and 4.04 THz for the LO and TO modes, respectively.
 The  recent Raman spectroscopy study of Olijnyk \etal \cite{Olijnyk06} puts the TO mode ($E_{2g}$) frequency at a somewhat higher value of 139 cm$^{-1}$. On the whole, the calculated frequencies are a bit lower than the INS results of
Wakabayashi \etal \cite{Wakabayashi71}  These authors have analyzed the INS results in terms of a sixth-neighbor modified axially symmetric force constant model and produce a frequency distribution, which reveals a maximum
frequency of about 7.25 THz or 241.8 cm$^{-1}$, approximately 10\% higher than the maximum calculated frequency 215 cm$^{-1}$ (Table \ref{table3}) in this work. The shape of the frequency distribution agrees in general with the phonon DOS
$F(\omega)$ shown in Fig. \ref{fig3}, except for the relative heights of some of the low frequency peaks.

The Raman spectroscopy work by Olijnyk \etal \cite{Olijnyk06} shows a shift of the $E_{2g}$ frequency under 
pressure from 139 cm$^{-1}$ to about 150 cm$^{-1}$ at 18.8 GPa. The  calculated $\Gamma$-point TO  
 mode frequency increases from 130.5 cm$^{-1}$ at the normal pressure lattice parameter ($a$= 6.255 a.u.) to 
 135.1 cm$^{-1}$ at  $a=$ 5.75 a.u., i.e., the increase is much more subdued. In Table \ref{table3} we have presented the calculated $E_{2g}$ frequency for various lattice parameters. Although the variation may seem non-monotonic, the differences are very much within the error bars of the calculation. What can perhaps be concluded is that there is no phonon-softening and most probably this frequency does increase by approximately 
5 cm$^{-1}$.  This is about 45-50\% of the observed shift. Note that calculations for the stable hcp phase for a lattice parameter between 5.75 and 5.6 a.u. might reveal a higher shift. The discrepancy between the calculated and the observed shift could be partly due to the possible variation in the $c/a$ ratio under pressure, which has not been considered in the calculation. 
Olijnyk \etal \cite{Olijnyk06} have analyzed the shift in the $E_{2g}$ frequency in terms of the relation
\begin{equation}
\label{E2g-c44}
\mu_{E_{2g}} = \frac{1}{2\pi}\left(\frac{4\sqrt{3}a^2C_{44}}{mc}\right)^{1/2}\;,
\end{equation}
where $m$ is the atomic mass and $C_{44}$ is the elastic shear modulus. They relate their result to the behavior of the shear modulus at the transition. The possible variation of the $c/a$ ratio with
pressure should be factored in in this discussion. The coarse wave vector mesh used for the phonon frequency calculation prevents us from using the slope of the phonon dispersion curves or the long wavelength method
to compute the elastic constants. {\it Ab initio} calculation of the elastic constants, including $C_{44}$,
via long wavelength and homogeneous deformation methods using energy-minimized $c/a$ ratios will be the subject of a separate publication.

It is noteworthy that both calculation and the Raman study show a positive shift of the $E_{2g}$ frequency, in contrast to the behavior in Y and the regular Lanthanides, where a softening of the $E_{2g}$ mode under
pressure is associated with the transition to the high-pressure Sm-type structure \cite{Olijnyk94,Olijnyk05}.
Incidentally, in our linear response calculation, as the lattice parameter is lowered to 5.6 a.u. where the instability of the hcp phase is manifest
via appearance of some imaginary frequencies, the $E_{2g}$ mode frequency stays real and has a lower value 
124 cm$^{-1}$.  At a  lattice parameter of 5.2 a.u., this frequency, along with many others,  are imaginary.

Broadening of the phonon bands and the increase in the maximum phonon frequencies with pressure is apparent in
Fig. \ref{fig3}. Maximum phonon frequency increases from 215 cm$^{-1}$ to  295 cm$^{-1}$, as the lattice 
parameter changes from the normal pressure value to 5.75 a.u., where the estimated pressure (Table \ref{table1})
is 22-23 GPa.  Note that with increasing pressure,
initial increases in $\bar{\omega}$, $\tilde{\omega}$ and $\omega_{\text{ln}}$ are followed by a decrease,
revealing phonon softening. At normal pressure, the shape of the $\alpha^2F(\omega)$ function follows that of
the phonon DOS $F(\omega)$. With increasing pressure, differences between the shape of the two functions appear due
to increased contribution from low frequency phonons, a consequence of phonon softening. There is a feed-back effect:  electron-phonon coupling leads to both phonon linewidth and renormalization of the phonon frequency, and the latter, in turn, affects the coupling constant.   Unfortunately, the 7,7,7
division of the BZ misses most of the symmetry points. However, it appears that there are several regions of
the wave vector space that show large mode coupling. Linear response calculations for the lattice parameter
5.6 a.u. show that the phonon frequency at the wave vector closest to the L-point for the 7,7,7 division
becomes imaginary first. For the lattice parameter 5.75 a.u., this wave vector has a very large mode coupling 
constant. 
In Table \ref{table3}, it is clear that general phonon softening starts as the lattice parameter increases
beyond 6.05 a.u. (as seen from $\omega_{\textit{ln}}$) or 5.85 a.u. (as seen from $\bar{\omega}$ and
$\tilde{\omega}$), while the $E_{2g}$ mode continues to stiffen and the maximum phonon frequency continues to increase.

The electron-phonon coupling parameter $\lambda_{ph}$ is a combination of an electronic parameter (Hopfield parameter)
$\eta = \langle I^2\rangle N(0)$ ($N(0)$ being the Fermi level DOS for one type of spin) and the mean square phonon frequency $\langle\omega^2\rangle$: $\lambda_{ph} = \eta/m\langle \omega^2\rangle$.
 $\langle I^2\rangle$ is the Fermi surface average of the square of the electron-phonon matrix element. The electronic  and phonon-related parameters act in opposite directions in affecting the coupling constant:  
 $\lambda$ is enhanced by having higher $\langle I^2\rangle$ at lower frequency. As the Hopfield parameter for transition metals has often been calculated using the rigid muffin-tin approximation
of Gaspari and Gyorffy \cite{RMT}, in Table \ref{table3} we have compared the values obtained  via the FP-LMTO
linear response method ($\eta$)  and those obtained by using the RMT scheme implemented within the 
LMTO-ASA method \cite{glotzel,skriver}, known as the rigid atomic sphere (RAS) method. The latter values, labeled as $\eta(RMT/RAS)$, are in general agreement with the FP-LMTO results, but are somewhat overestimated, with the level of the overestimation increasing with pressure. This is in contrast with the results
for the late transition metal hcp Fe, where RAS was shown to consistently underestimate the Hopfield parameter,
with the level of the underestimation increasing with pressure \cite{Bose03}. The values of the
Fermi surface averaged electron-phonon matrix element $\langle I^2\rangle$, obtained from the linear response calculation and by using the RMT approximation, are also shown in Table \ref{table3}.
\begin{table*}
\caption{
Hopfield parameters from the linear
response calculation $\eta$ and the rigid muffin-tin (atomic sphere) approximation $\eta$ (RMT/RAS), the
Fermi surface average of the square of the electron-phonon matrix element $\langle I^2\rangle$  and
$\langle I^2\rangle$(RMT/RAS) from the linear response and the RMT results, respectively;
 logarithmic
average phonon frequencies $\omega_{\ln}$;  maximum phonon frequencies $\omega_m$, average phonon frequencies
$\bar{\omega}=\langle \omega\rangle$ and $\tilde{\omega}=\langle \omega^2\rangle^{1/2}$; the $\Gamma$-point
$E_{2g}$ mode frequency $\omega(E_{2g})$;
Coulomb pseudopotentials for Eliashberg equation ($\mu^*(\omega_c)$)
and McMillan formula ($\mu^*_{\text{ln}}$);
electron-phonon coupling parameters $\lambda_{ph}$;
 calculated critical temperatures
($T_c^{\text{calc}}$) 
from the solution of the Eliashberg equations (\ref{eliash})
and the critical temperatures  from the McMillan formula (\ref{mcm})($T_c^{\text{McM}}$)
for various lattice parameters $a$.
}
\label{table3}
\begin{ruledtabular}
\begin{tabular}{lrcccccc}
$a$ & $a_{\text B}$  & 6.255   & 6.15   & 6.05   & 5.85   & 5.75\\   
$\eta$& Ry/bohr$^2$  & 0.0381 & 0.0433 & 0.0519 & 0.0666 & 0.0771\\  
$\langle I^2\rangle$ & (Ry/bohr)$^2$ & 0.0026 & 0.0031 & 0.0040 & 0.0056 & 0.0065\\
$\eta$ (RMT/RAS)& Ry/bohr$^2$
                     & 0.0439 & 0.0509 & 0.0591 & 0.0826 & 0.1077\\ 
$\langle I^2\rangle$(RMT/RAS) & (Ry/bohr)$^2$ & 0.0030 & 0.0037 & 0.0045 & 0.0065 & 0.0081\\
\hline
     &     K         & 160.4   & 195.0   & 198.1   & 195.4   & 177.7\\   
\rb{$\omega_{\text{ln}}$} & cm$^{-1}$
                     & 111.5   & 135.5   & 137.7   & 135.8   & 123.5\\  
\hline
$\omega_m$&cm$^{-1}$ & 214.5  & 230.4  & 246.9  & 279.6  & 294.7\\ 
$\omega(E_{2g})$ &cm$^{-1}$ & 130.5 & 134.95 & 133.4 & 136.14 & 135.0 \\ 
$\bar{\omega}$&cm$^{-1}$ &122.4 & 142.3 & 145.1 & 145.6 & 135.3\\ 
$\tilde{\omega}$&cm$^{-1}$ &131.5 & 148.6 & 152.3 & 155.8 & 148.1\\
$\mu^*(\omega_c)$&   & 0.253  & 0.256  & 0.260  & 0.266  & 0.269 \\  
$\mu^*(\omega_{\text{ln }})$&
                     & 0.160  & 0.161  & 0.163  & 0.165  & 0.166 \\  
$\lambda_{ph}$  &    & 0.639 & 0.576  & 0.657 & 0.807 &  1.033\\
$T_c^{\text{McM}}$& K&2.34  & 1.66  & 2.96 & 5.66  & 9.18\\
$T_c^{\text{calc}}$& K& 2.17 & 1.75  & 3.13 & 6.00  & 9.64\\  
\end{tabular}
\end{ruledtabular}
\end{table*}

\section {Superconductivity in hcp S\lowercase{c}, based purely on electron-phonon coupling}

Since {\it ab initio} results of the electron-phonon coupling in hcp Sc have not been reported
in the literature, it would be of some interest to derive values of the superconducting
transition temperature $T_c$ based on the present calculations. Indeed, the values of
the coupling constant $\lambda_{ph}$ listed in Table \ref{table3} suggest that $T_c$
could be high enough to be experimentally observable. The superconducting transition 
temperature can be obtained by solving the
linearized isotropic Eliashberg equation at $T_c$(see, e.g., Ref. \onlinecite{allen-mitro}):
\begin{eqnarray}
Z(i\omega _{n}) &=&1+\frac{\pi T_{c}}{\omega _{n}}\sum_{n^{\prime
}}W_{+}(n-n^{\prime })\mbox{sign}(\omega_{n^{\prime }}),  \label{eliash} \\
Z(i\omega _{n})\Delta (i\omega _{n}) &=&\pi T_{c}\sum_{n^{\prime }}^{\left|
\omega _{n}\right| \ll \omega _{c}}W_{-}(n-n^{\prime })\frac{\Delta (i\omega
_{n^{\prime }})}{\left| \omega _{n^{\prime }}\right| },  \notag
\end{eqnarray}%
where $\omega _{n}=\pi T_{c}(2n+1)$ is a Matsubara frequency,
$\Delta (i\omega _{n})$ is an order parameter and $Z(i\omega _{n})$ is
a renormalization factor.  
Interactions $W_{+}$ and $W_{-}$ contain a
phonon contribution $\lambda _{ph}$, a contribution from spin fluctuations $\lambda _{sf}$,
and effects of scattering from impurities.  With scattering rates $\gamma _{m} = \frac{1}{2\tau_m}$
and $\gamma _{nm} = \frac{1}{2\tau_{nm}}$ referring to magnetic
and nonmagnetic impurities, respectively, the expressions for the interaction terms are :
\begin{eqnarray}
W_{+}(n-n^{\prime })&=&\lambda _{ph}(n-n^{\prime })+\lambda _{sf}(n-n^{\prime
})\notag \\
& & +\delta _{nn^{\prime }}(\gamma _{nm}+\gamma _{m}), 
\label{eliash2}
\end{eqnarray}
and
\begin{eqnarray}
W_{-}(n-n^{\prime })&=&\lambda _{ph}(n-n^{\prime })-\lambda _{sf}(n-n^{\prime
}) \notag \\
& & -\mu ^{\ast }(\omega _{c})+\delta _{nn^{\prime }}(\gamma _{nm}-\gamma _{m}).
\label{eliash3}
\end{eqnarray}
The phonon contribution is given by
\begin{equation}
\lambda _{ph}(n-n^{\prime })=2\int_{0}^{\infty }\frac{d\omega \omega\alpha
^{2}(\omega )F(\omega )}{(\omega _{n}-\omega _{n^{\prime }})^{2}+\omega ^{2}}%
\;,
\end{equation}%
where $\alpha ^{2}(\omega )F(\omega )$ is the  Eliashberg spectral
function, defined as
\begin{equation}
\alpha^2F(\omega) =\frac{1}{N(0)}\sum_{{\bf k},{\bf k'},ij,\nu} |g_{{\bf k},{\bf k'}}^{ij,\nu}|^2
\delta(\varepsilon_{{\bf k}}^i) \delta(\varepsilon_{{\bf k'}}^j) \delta(\omega - 
\omega_{{\bf k}-{\bf k'}}^\nu)\;.
\end{equation}
 Here,  
 $g_{{\bf k},{\bf k'}}^{ij,\nu}$ is 
the electron-phonon matrix element, with $\nu$
being the phonon polarization index and ${\bf k},{\bf k'}$ representing
electron wave vectors with band indices $i$, and $j$, respectively.
$\lambda_{ph}(0) =\lambda_{ph}$ is the electron-phonon coupling parameter, the values of which
are given in Table \ref{table3}.  
The contribution connected  with spin fluctuation can be written as
\begin{equation}
\label{lambdasf}
\lambda _{sf}(n-n^{\prime })=\int_{0}^{\infty }\frac{d\omega ^{2}P(\omega )}{%
(\omega _{n}-\omega _{n^{\prime }})^{2}+\omega ^{2}}\;,
\end{equation}%
where
 $P(\omega )$  is  the spectral function of spin fluctuations, related
to the imaginary part of the  transversal spin susceptibility $\chi _{\pm }(\omega )$ as
\begin{equation*}
P(\omega )=-\frac{1}{\pi }\left\langle \left| g_{\mathbf{kk}^{\prime
}}\right|^{2}\mathrm{Im}\chi _{\pm }(\mathbf{k,k}^{\prime },\omega
)\right\rangle_{FS}\;,
\end{equation*}
where
$\langle \;\;    \rangle_{FS}$ denotes Fermi surface average. $\lambda _{sf}= \lambda _{sf}(0)$ is
often referred to as the electron-paramagnon coupling constant.

In Eq.(\ref{eliash3}), $\mu ^{\ast }(\omega _{c})$ is the  screened Coulomb interaction, 
\begin{equation}
\mu ^{\ast }(\omega _{c})=\frac{\mu }{1+\mu \ln (E/\omega _{c})},  \label{mu}
\end{equation}%
 with $\mu =\left\langle N(0)V_{c}\right\rangle _{FS}$ being the Fermi surface average
 of the
Coulomb interaction.  E is a characteristic electron energy, usually chosen as the Fermi energy
$E_F$ and $\omega
_{c}$  is a cut-off frequency, usually chosen ten times the
maximum phonon frequency: $\omega
_{c} \simeq 10\omega _{ph}^{\max }$.

For a start, we ignore all consideration of spin fluctuations and impurity scattering and solve the Eliashberg
equation with only the electron-phonon term and the Coulomb pseudopotential $\mu^{\ast}(\omega_{c})$. 
  As is  often done, we assume that a reasonable value for  $\mu$ is $\sim 1.0$, and 
   from the calculated Fermi energies $E_F$  we obtain
   $\mu ^{\ast }$ for all volumes, with the cut-off frequency $\omega_{c}$ assumed to be
   ten times the maximum phonon frequency. The values are listed in Table \ref{table3}.
As shown in this table, the superconducting transition temperature in hcp Sc based on consideration of electron-phonon coupling alone can be significant, increasing monotonically with pressure from 2K to $\sim 10$K
until the instability of the hcp phase sets in. Interestingly, the $T_c$ values 
reported by Hamlin and Schilling \cite{Schilling07} for the high pressure Sc-II phase fall in this range, while
no superconducting behavior has been observed in the hcp phase.

For pedagogical reasons, we have listed in Table\ref{table3} the values of $T_c$ obtained by using the
  Allen-Dynes form\cite{allen-mitro} of the
 McMillan expression:
    \begin{equation}
       T_{c}=\frac{\omega_{ln}}{1.2}\exp \left\{ -\frac{1.04 \left(1+\lambda_{ph} \right)}{%
  \lambda -\mu ^{\ast }(1+0.62\lambda_{ph})}\right\} ,  \label{mcm}
  \end{equation}
 where $\omega_{ln}$ is the logarithmically averaged phonon frequency\cite{allen-mitro}, obtained
 from our linear response calculations and reported in Table \ref{table3}. Note that the Coulomb pseudopotential $\mu^{\ast}$ appearing
 in the McMillan equation above is related to $\mu ^{\ast }(\omega_{c})$ appearing in the
 Eliashberg equation via\cite{allen-mitro}
 \begin{equation}
   \mu^{\ast} = \mu^{\ast}\left(\omega_{ln}\right) = \frac{\mu ^{\ast }(\omega_{c})}
   {\left(1+\mu ^{\ast }(\omega_{c})\ln\left(\omega_{c}/\omega_{m}\right)\right)}\;.
   \end{equation}
  Our results are computed with $\omega_{c}/\omega_{m}
 =10$.  The $T_c$ values obtained by solving the Eliashberg equations and those from the McMillan expression
 Eq.(\ref{mcm}) show excellent agreement. Earlier, calculations for hcp Fe \cite{Bose03} had shown the McMillan expression to overestimate $T_c$ with respect to the results from the Eliashberg equation, while for fcc and bct boron an opposite trend was revealed \cite{Bose05}.

\section  {Spin fluctuation effects}
\subsection{hcp Sc at normal pressure}

Faced with the results of the previous section, one needs to produce convincing arguments as to why hcp Sc is not
 superconducting despite sufficiently strong electron-phonon coupling. A  mechanism that is known to cause  suppression of superconductivity is spin fluctuations, which is often quoted as the reason why fcc Pd is not
superconducting\cite{Berk-Schrieffer}. It is argued that the high value of $N(0)$ in fcc Pd leads to considerable Stoner enhancement of paramagnetic spin susceptibility, making it a borderline ferromagnetic material, and spin fluctuations tending to
ferromagnetic alignment of spins lead to the breaking of the Cooper pairs. That such a mechanism is operative in hcp Sc as well is highly probable and has been discussed in the literature on a few occasions\cite{Capellmann,Das}.
Jensen and Maita\cite{Jensen-Maita} argue that the spin fluctuation effects are responsible for the rapid depression of $T_c$ in the Zr-Sc alloy system, as Sc is added. {\it Ab initio} calculations of the spin susceptibility of hcp Sc at equilibrium volume based on the spin-density functional theory by MacDonald\etal \cite{MacDonald77} yield an exchange-correlation enhancement factor of 4.03 over the band value and 17.2 over the free-electron value at the same average electron density as in hcp Sc. An earlier calculation by Das\cite{Das} puts the Stoner enhancement factor at 4.6 (over the band value). The value calculated for fcc Pd by  Janak\cite{Janak} is 4.46. This shows that the spin fluctuations in hcp Sc should be as strong as in fcc Pd.

For a proper theoretical treatment of the spin fluctuation effects one needs to compute $\lambda _{sf}(n-n^{\prime })$ from the
spin susceptibility function given by Eq.(\ref{lambdasf}). However, it is important to note that such treatments tacitly assume a Migdal-like theorem being applicable to spin fluctuations. The Eliashberg equations
(Eq.(\ref{eliash})) are based on the assumption that the maximum or the cut-off energy of spin
fluctuation is much smaller than the characteristic electronic energy, e.g. the Fermi level. A somewhat qualitative treatment of spin fluctuations can be based on
estimating $\lambda_{sf} =\lambda_{sf}(0)$ from experiments. 
Both electron-phonon and the electron-paramagnon interactions contribute to the electronic specific heat. In an independent one-electron picture this is interpreted as the electronic mass enhancement or equivalently,
enhancement of the density of states over the bare value $N(0)$.  The latter is the value given by calculations, where these interactions are not included in the one-electron Hamiltonian.  Thus, a  reliable estimate of the electron-paramagnon coupling constant $\lambda_{sf}$ can be obtained from the measured value of the temperature co-efficient of the electronic specific heat $\gamma$, and the calculated values of the bare electron density of states and the electron-phonon coupling constant $\lambda_{ph}$:
\begin{eqnarray}
\gamma & = & \frac{\pi^2}{3}k_{B}^2 N^{\ast}(0)\\
N^{\ast}(0) & = & N(0)(1+ \lambda_{eff})\\
\lambda_{eff} & = & \lambda_{ph}+ \lambda_{sf}\;.
\end{eqnarray}
Here, $\gamma$ and $N(0)$ refer to the values per atom. The Coulomb interactions are included in an average sense in the density functional calculations of $N(0)$, and have therefore been left out of Eq. (12).

Among all the elemental metals, excluding the rare-earths, Sc has the largest electronic specific heat constant $\gamma$, followed by Y and Pd\cite{Gschneidner}. The reported experimental values at normal pressure are 10.9-10.33 mJ/mole K$^2$\cite{Jensen-Maita,Gschneidner,Tsang,Reichardt,Flotow,Swenson}. Considering the latest and the most conservative value of $\gamma=$10.33 mJ/mole K$^2$\cite{Tsang,Reichardt}, we get $N^{\ast}(0)$= 2.2 states/(eV atom spin). With the calculated value of 1.067 states/(eV atom spin), we get $\lambda_{eff}$ = 1.063.
The calculated value of $\lambda_{ph}=$ 0.639 then yields $\lambda_{sf}=$0.422. Effects of spin fluctuations on $T_c$ can be incorporated by a simple rescaling of $\lambda_{ph}$ and the Coulomb pseudopotential $\mu^{\ast}$:
$\lambda_{ph}\rightarrow\lambda_{ph}/(1+\lambda_{sf})$, $\mu^{\ast}\rightarrow 
(\mu^{\ast}+\lambda_{sf})/(1+\lambda_{sf})$\cite{Mitrovic}. An extension of the McMillan
formula\cite{mazin1} that is often used to incorporate the spin fluctuation effects is
\begin{equation}
T_{c}=\frac{\omega _{\ln  }^{ph}}{1.2}\exp \left\{ -\frac{1.04(1+\lambda
_{ph}+\lambda _{sf})}{\lambda _{ph}-\lambda _{sf}-\mu ^{\ast
}[1+0.62(\lambda _{ph}+\lambda _{sf})]}\right\} .  \label{maz}
\end{equation}%
This formula is meaningful as long as $\lambda _{sf}$ is sufficiently less than $\lambda _{ph}$, so that the 
denominator in the argument of the exponential in Eq. (\ref{maz}) stays positive and not close to zero. For the above values of  $\lambda_{ph}$, $\lambda_{sf}$ and the Coulomb pseudopotential $\mu ^{\ast}$=0.16 (see Table
\ref{table3}), this condition breaks down. In fact, for comparable values of $\lambda _{ph}$ and $\lambda_{sf}$,
the Kernel of the Eliashberg equation (Eq. (\ref{eliash})) $W_{-}(n-n^{\prime })$, given by Eq.(\ref{eliash3}), becomes negative, allowing no solution for $T_c$. It should be emphasized that the McMillan-type
formulas in the presence of spin fluctuations are good for rough ball-park estimates of $T_c$ only. 

Note that our calculated value of $\lambda_{ph}=0.639 $ for normal pressure hcp Sc is significantly higher than the value $\lambda_{ph}=0.3$ that has been suggested by Knapp and Jones\cite{Knapp}, based on a comparison of the high and low temperature electronic specific heats. A value of $\lambda_{ph}=0.3$ would suggest $\lambda_{sf} =0.76$, more than twice larger than  $\lambda_{ph}$. Our results suggest that $\lambda_{sf} < \lambda_{ph}$, but their values are close. This leads to the possibility that if spin fluctuations can be suppressed via application of pressure, superconductivity can indeed appear. 

Based on the analysis of high and low temperature specific heats, Knapp and Jones\cite{Knapp} also suggested that for Pd $\lambda_{ph}=0.7$. The electronic specific heat constant $\gamma$ for Pd is only slightly lower than that in hcp Sc\cite{Gschneidner,Knapp}, lying between 10.0 and 9.2 mJ/mole K$^2$. The calculated bare band density of states $N(0)$ for fcc Pd is about 1.3 states/(eV atom spin), giving $\lambda_{eff}$ in the range 0.66 and 0.5.
An analysis by Savrasov and Savrasov\cite{savrasov2} puts the value at 0.69. The value $\lambda_{ph}=0.7$ suggested by Knapp and Jones would leave no room for spin fluctuation effects in Pd, and would render Pd superconducting. The FP-LMTO linear response
calculation for fcc Pd by  Savrasov and Savrasov\cite{savrasov2} yields $\lambda_{ph}=0.35$, giving $\lambda_{sf}$ in the range 0.3-0.15. These values of $\lambda_{ph}$ and $\lambda_{sf}$ can adequately explain the
nonexistence of superconductivity in fcc Pd. We have repeated the linear response calculation for fcc Pd and
obtain a value of $\lambda_{ph}=0.37$, in close agreement with the result of Savrasov and Savrasov\cite{savrasov2}. Note that the  calculation by Pinski and Butler\cite{Pinski} based on the RMT approximation yields $\lambda_{ph}=0.41$ for fcc Pd, in reasonable agreement with the FP-LMTO linear response
results. To summarize, our linear response calculations and electronic specific heat analysis show that in both
hcp Sc and fcc Pd the values of electron-phonon  and electron-paramagnon coupling constants are comparable, with the latter being slightly lower. This is qualitatively different from the previous results of Knapp and Jones\cite{Knapp}, which would suggest that in Sc $\lambda_{ph}$ is significantly lower than $\lambda_{sf}$, 
and {\it vice versa} for fcc Pd. Note that according to our results both $\lambda_{ph}$ and $\lambda_{sf}$
are larger in hcp Sc than the corresponding quantities in fcc Pd.

While the nature of spin fluctuations in fcc Pd is accepted to be ferromagnetic, that in hcp Sc could be
antiferromagnetic. The dynamic susceptibility $\chi(q,\omega)$ in Pd has a peak at $q=0$, whereas $\chi(q,\omega)$
in hcp Sc is expected to have a peak at some finite wave vector $Q$. First-principles calculation of 
$\chi(q)$ including spin-orbit coupling and all other relativistic effects by Thakor\etal\cite{Thakor} yields
a peak at the wave vector ${\bf q}=(0,0.0.57)\pi/c)$, consistent with a Fermi surface nesting vector.
Earlier calculations produce a peak along the same direction but at different values\cite{Rath,Liu} of
$q$. Capellmann\cite{Capellmann} has discussed the effect of incipient antiferromagnetism in Sc and shown that
the antiferromagnetic spin fluctuations should lead to a repulsive electron-electron interaction, resulting
in suppression of superconductivity. The modified McMillan formula (Eq.(\ref{maz})) has been used to estimate
$T_c$ for both ferromagnetic and antiferromagnetic spin fluctuations\cite{mazin1,Bose03}.

\subsection{Superconductivity in hcp Sc at high pressure immediately before the transition to Sc-II phase}

Our linear response calculation yields a high value $\lambda_{ph}\sim 1.0$ of the electron-phonon coupling 
constant at high pressure hcp phase prior to the transition to the complex Sc-II structure. It is natural
to ask whether spin fluctuations should be able to suppress superconductivity despite such strong
electron-phonon coupling. Since specific heat data are not available for such high pressures, a reliable estimate
of $\lambda_{sf}$ is difficult. However, a reasonable step might be to scale the ambient pressure $\lambda_{sf}$ according to the density of states $N(0)$ at high pressure shown in Table \ref{table1}. This gives $\lambda_{sf}=0.346$ for the smallest volume or the highest pressure calculation. With $\lambda_{ph}=1.0,
\lambda_{sf}=0.346, \mu^{\ast}=0.166$, and $\omega_{ln}=177.7 K$, as shown in Table \ref{table3}, and using 
Eq.(\ref{maz}), we get $T_c$ = 0.14 K. As pointed out earlier, at comparable values of $\lambda_{ph}$ and $\lambda_{sf}$, Eq.(\ref{maz}) may not be reliable, and one needs to solve the Eliashberg equation. However, our estimate of $T_c=.14 K$ indicates that there is a possibility of hcp Sc turning superconducting just before the onset of the high pressure Sc-II phase. This is indeed what has been reported by Wittig \etal \cite{Wittig79}.
Note that such a conclusion would not be tenable for smaller values of $\lambda_{ph}$, as the results of 
Knapp and Jones\cite{Knapp} would suggest.

 The use of energy-optimized $c/a$ ratio at higher than ambient pressures may change the calculated values of $\lambda_{ph}$ somewhat, but such small changes are not expected to influence our results qualitatively and   will not affect the nature of the conclusions. 

\subsection{Superconductivity in Sc-II phase}

Spin fluctuations are expected to be significantly reduced in the Sc-II phase. According to the calculations of
Ormeci \etal \cite{Ormeci06},  the density of states at the Fermi level $N(0)$ in the Sc-II structure should be
about half the value in the ambient pressure hcp Sc. This suggests that $\lambda_{sf}$ should be reduced to a
value of 0.2 or less.  Reasonable choices of $\omega_{ln}$ in the Sc-II phase at pressures $\sim$30 GPa (see Table \ref{table3}) and $\mu^{\ast}$ should be 210.0 K and 0.166, respectively. A value of $\lambda_{ph}$=0.95
in Eq.(\ref{maz}) would yield a $T_c$=1.4 K. Reducing the value of $\lambda_{sf}$ to 0.1 would result in a
$T_c$ of 4.5 K. These numbers suggest that superconductivity in the Sc-II phase can be explained based on the
quantities calculated for the high pressure hcp Sc and reasonable estimates of the reduction in the spin
fluctuation effects that is expected as a result of the change in the density of states. Note that phonon softening causes $\lambda_{ph}$ to increase in the hcp phase immediately before the transition to the Sc-II
structure. At the start of the Sc-II phase the value of $\lambda_{ph}$ should be somewhat less than the value 
1.033 shown in Table \ref{table3}, hence the choice 0.95. It should be noted that although the density of states
$N(0)$ decreases with increasing pressure, there is a rapid increase in the electron-phonon matrix element
$\langle I^2\rangle$ with decreasing volume (see Table IV). A high value of $\lambda_{ph}$ in the Sc-II phase is thus possible despite the reduced value of $N(0)$. Chances are that at the start of the Sc-II phase both $\lambda_{ph}$ and $\lambda_{sf}$ are somewhat less than the values suggested here.

\section {Comparison between hcp S\lowercase{c} and
 hcp F\lowercase{e}}

The bulk moduli of all transition metals are known to increase (at least initially) as a function of band 
filling \cite{varenna2}, and the bulk moduli of the late transition metals are in general higher than those
of the early ones.  The bulk modulus of Fe is thus expected to be larger than that of Sc, a consequence of Fe having more $d$-electrons.  The calculations of Bose \etal \cite{Bose03} for the hcp phase of Fe show values
in the range of 300 GPa to 970 GPa, corresponding to volumes per atom for which the calculated $T_c$ is above zero
(the theoretical superconducting phase). These values are 4-5 times higher than the bulk modulus values of Sc
shown in Table \ref{table1}. Accordingly, the average phonon frequency in hcp Fe is about four times higher than
in Sc.  Because of a much larger
number of $d$-electrons  in hcp Fe (6.5 on average) than Sc (1.6 on average), the electron-phonon matrix element
$\langle I^2\rangle$ in Fe is also much larger. As a result, despite the lower value of $N(0)$, the Hopfield parameter $\eta = N(0)\langle I^2\rangle$  in hcp Fe turns out to be more than double that in Sc (compare Table \ref{table3} with TABLE II of Ref. 
[\onlinecite{Bose03}]). However, the dominating effect turns out to be the lower phonon frequencies in Sc,
leading to higher values of $\lambda_{ph}$. 

Spin fluctuations, most probably of an antiferromagnetic nature\cite{Bose03,mazin1}, are believed to be present in the high  pressure hcp phase of Fe. Because the density of states $N(0)$ in hcp Fe is about half that in hcp Sc, spin fluctuations in hcp Fe should be much weaker. There is no specific heat data available for hcp Fe, the stable
phase of Fe at pressures of $\sim$10 GPa and higher. As a result, an experimental estimate of $\lambda_{sf}$ in hcp Fe is not available. Some theoretical estimates were provided in Refs. [\onlinecite{Bose03,mazin1}].

For a crude estimate of the ratio between the values of $\lambda_{ph}$ for hcp Fe and Sc, we can assume the Hopfield
parameter $\eta$ to be simply proportional to the fractional $d$-DOS, $N_d(0)/N(0)$. For an estimate of $m\langle \omega^2\rangle$, we can use the result proposed by Moruzzi, Janak and Schwarz \cite{MJS}, relating the Debye
temperature $\Theta_D$ to the bulk modulus, atomic mass and the average Wigner-Seitz radius:
\begin{equation}
\label{mjs}
\Theta_D = 41.63\sqrt{\frac{s_0 B}{m}}\;,
\end{equation}
where $s_0$ is the average Wigner-Seitz radius in a.u. and $B$ is the bulk modulus in kbar. This result suggests that $m\langle\omega^2\rangle
\propto s_0 B $. Using $B$ values from Table \ref{table1} and TABLE  I of [\onlinecite{Bose03}], the DOS values
from Table \ref{table2}, $s_0$ computed from $a= 4.6$ a.u, $c/a=\sqrt{8/3}$ for Fe and $a=6.255$ a.u., $c/a=
1.592$ for Sc, we get $\lambda_{Fe}/\lambda_{Sc} = 0.45$. This compares favorably with the value 0.66, according  to the computed linear response values (Table \ref{table3} and TABLE II of [\onlinecite{Bose03}]). We have
used $N_d(0)=34.55$ states/(Ry cell) and $N(0)=40.79$ states/(Ry cell) (TABLE I of [\onlinecite{Bose03}]) for Fe.
The proportionality of $\eta$ to fractional $d$-DOS, $N_d(0)/N(0)$, can be somewhat justified on the basis of the RMT result of Gaspari and Gyorffy \cite{RMT}:
\begin{equation}
\eta = 2N(0)\sum_l (l+1)M^2_{l,l+1}\frac{f_l}{2l+1}\frac{f_{l+1}}{2l+3}\;,
\end{equation}
where $N(0)$ is the Fermi level DOS per atom per spin and $f_l$ is a relative partial state density,
\begin{equation}
f_l = \frac{N_l\left(0\right)}{N(0)}\;.
\end{equation}
 $M_{l,l+1}$ is the electron-phonon matrix element obtained from the gradient of the potential and the
radial solutions $R_l$ and $R_{l+1}$ of the Schr\"{o}dinger equation evaluated at the Fermi energy.
 Neglecting the matrix elements and all partial DOSs other than $N_d(0)$ amounts to the result
 $\eta \propto N_d(0)/N(0)$. Of course, this provides a very crude estimate and can only be relied on in
deciding whether $\lambda$ for one material is greater or lower than for the other and no quantitative
reliability can be guaranteed. In fact, in getting the estimate $\lambda_{ph,Fe}/\lambda_{ph,Sc} = 0.45$ we have 
ignored the possible differences in the $p$- and $f$- DOSs of the two solids. Consideration of these
differences would lead to an inferior quantitative agreement. Also, the ratio $m\langle\omega^2\rangle$ for Fe to
Sc for the volumes considered is 2.35 according to Eq.(\ref{mjs}), while the value from the
linear response calculation is 3.5.

\section {Summary}

FP-LMTO linear response calculations for the hcp phase of Sc, based on a fixed $c/a$ ratio of 1.592, shows a monotonic increase in electron-phonon coupling with pressure. Calculated phonon frequencies for the equilibrium lattice parameter are 6-10\% lower than those obtained via INS experiments\cite{Wakabayashi71}. The agreement  can perhaps be improved with calculations done on a finer wave vector mesh.  The estimated pressure, based on the linear response results, for the instability to appear is 23-30 GPa. This pressure range,
which can certainly be narrowed with additional calculations, agrees with the experimental observations. The $\Gamma$-point $E_{2g}$ frequency shows a modest increase with pressure, in qualitative agreement with the
Raman work by Olijnyk \etal \cite{Olijnyk06}  Energy-optimized choice of the $c/a$ ratio for each volume per atom may lead to a better agreement.

The electron-phonon coupling constant $\lambda_{ph}$ is found to
increase steadily with pressure in the hcp phase, until the pressure reaches a value where the hcp phase becomes unstable. An estimate of the electron-paramagnon coupling constant based on the measured temperature coefficient of the electronic specific heat, calculated band density of states and $\lambda_{ph}$ suggests that the spin 
fluctuations at normal pressure should be strong enough to suppress superconductivity completely. At the
highest pressures where the hcp phase is still stable, the increase in $\lambda_{ph}$ and a decrease in 
$\lambda_{sf}$ in proportion to the calculated band density of states suggest the possibility of a very low $T_c$
superconductivity, as noted by Wittig \etal \cite{Wittig79} A comparison of the band densities of states in the
hcp and the Sc-II phases shows that the spin fluctuation effects in the Sc-II phase should be reduced by a factor of two or more. It is argued that this suppression of spin fluctuation combined with electron-phonon coupling
constants of a magnitude similar to that calculated for the high pressure hcp phase can indeed account for the
observed superconductivity in the Sc-II phase.

ACKNOWLEDGMENTS

This work was supported by a grant from the Natural Sciences and Engineering Research Council of Canada. The
author acknowledges helpful discussions with B. Mitrovi\'{c}, J. Kortus and O. Jepsen.

\end{document}